\documentclass[preprint,tightenlines, nofootinbib,aps,prc]{revtex4-1} 

\usepackage{graphicx}
\usepackage{bm}
\usepackage{url,multirow,xcolor}

\begin{document}

\title{ Consistency Examinations of Calculations of Nuclear Matrix Elements of Double-$\bm{\beta}$ Decay by QRPA }\thanks{Presented in the 37th International Workshop on Nuclear Theory, June 24$-$June 30, 2018, the Rila Mountains, Bulgaria.}

\author{J.\ Terasaki}
\affiliation{ Institute of Experimental and Applied Physics, Czech Technical University in Prague, Horsk\'a 3a/22, 128 00, Prague 2, Czech Republic \\ }

\begin{abstract} 
The neutrinoless double-$\beta$ decay is a hypothetical rare nuclear decay, which can be used for determining the neutrino-mass scale. The scheme to use this decay for determining the neutrino-mass scale is one of few limited methods possible to determine that. Nuclear matrix element of this decay is an important input to this method, and this matrix element cannot be determined by experiment. I examine the validity of the transition density used for calculating the nuclear matrix element by comparing the experimental data and my calculated result of the charge-change strength functions of $^{48}$Ca and $^{48}$Ti. The nuclear wave functions are obtained by the quasiparticle random-phase approximation. A new idea is proposed on the transition operator for this strength function, and the data of those nuclei are  reproduced well consistently. Reduced half-life of a few nuclei to the neutrinoless double-$\beta$ decay are shown. \end{abstract}

%
\maketitle
\section{Introduction}
 Since the discovery of the neutrino oscillation \cite{Fuk98,Ahm02,Egu03,Ali05}, the determination of the neutrino mass is one of the most important subjects of modern physics. The finite neutrino mass implies the necessity of extension of the standard theory. The neutrino mass may affect the mass distribution in the universe because of the abundance of the neutrino, although the neutrino mass is extremely small. The methods possible to determine the neutrino mass are quite limited, and one of the methods is to use the neutrinoless double-$\beta$ decay. If the neutrino is a Majorana particle, the effective neutrino mass can be determined by the half-life of the decay expected to be measured by the experiments and the transition matrix elements obtained theoretically. 

The nuclear part of the transition matrix element, called nuclear matrix element, is difficult to establish, because the nuclear wave functions are necessary for the medium to heavy nuclei for which approximation is essential. The nuclear matrix elements calculated by several methods and different groups are distributed to a range of factor 2$-$3 \cite{Eng17}. Reduction of this uncertainty factor is the most urgent task for the nuclear theory because of the necessity of the accurate determination of the effective neutrino mass and the design of the future experimental setup for detecting the neutrinoless double-$\beta$ decay. 
The nuclear matrix element cannot be confirmed experimentally because the effective neutrino mass is unknown. Thus, the subject for the nuclear theorists is to improve the reliability of the calculated nuclear matrix elements by examining and improving the calculations.  

In this paper, I check the charge-change transition density of my calculation indirectly by comparing the calculated and experimental charge-change strength functions for $^{48}$Ca $\rightarrow$ $^{48}$Sc and  $^{48}$Ti $\rightarrow$ $^{48}$Sc \cite{Yak09}. This check is important, considering the difficulty of the confirmation of the nuclear matrix elements mentioned above. It is, however, not trivial to reproduce the experimental charge-change strength function because the data do not satisfy the Gamow-Teller sum rule. The main new point of this paper is to clarify what those experimental data tell us. The transition density is checked by addressing this issue. 

\section{Scheme to determine effective neutrino mass using double-$\bm{\beta}$ decay}
The scheme is well established these days, e.g.~\cite{Doi85}. The half-life of the neutrinoless double-$\beta$ decay $T^{(0\nu)}_{1/2}$, nuclear matrix element $M^{(0\nu)}$, phase-space factor  $G^{(0\nu)}$ arising from the emitted electrons, and the effective neutrino mass $\langle m_\nu \rangle$
have the relation
\begin{eqnarray}
\frac{1}{ T^{(0v)}_{1/2} } = g_A^4 G^{(0\nu)}| M^{(0\nu)}|^2 \left( \frac{\langle m_\nu \rangle}{m_e} \right)^2, \label{eq:transition_probability_0vbb}
\end{eqnarray}
where $g_A$ denotes the axial-vector current coupling, and $m_e$ is the electron mass. The effective neutrino mass is defined by 
\begin{eqnarray}
\langle m_\nu \rangle = \left| \sum_{i=1,2,3} U_{ei}^2 m_i \right|.
\end{eqnarray}
$U_{ei}$ is the matrix element of Pontecorvo-Maki-Nakagawa-Sakata (PMNS) matrix \cite{Pon58} with $i$ denoting the mass eigen state ($e$ stands for the electron flavor). The information on the PMNS matrix elements is exploited rapidly recently by many neutrino-oscillation experiments leaving only few unknown parameters \cite{Neu18}. However, each eigen mass cannot be determined by those experiments, therefore, $\langle m_\nu \rangle$ is unknown. If $T^{(0\nu)}$ is obtained by the experiments of the neutrinoless double-$\beta$ decay, and $M^{(0\nu)}$ and $G^{(0\nu)}$ are calculated reliably, $\langle m_\nu \rangle$ can be determined. If $U_{ei}$ is complex, $\langle m_\nu \rangle$ would be different from the expectation value of the electron-neutrino mass. In any case, $\langle m_\nu \rangle$ is thought to be a neutrino-mass scale. In this paper, I always consider the ground-state-to-ground-state decay, and the value of $G^{(0\nu)}$ is cited from Ref.~\cite{Kot12} for my calculations. $M^{(0\nu)}$ in the  quasiparticle random-phase approximation (QRPA) \cite{Rin80} is calculated according to the equation 
\begin{eqnarray}
M^{(0\nu)} = \sum_{b_i b_f} \sum_{pp^\prime} \sum_{nn^\prime}
\langle pp^\prime | V(r_{12},\bar{E}_b) | nn^\prime\rangle
\langle 0^+_f | c_{p^\prime}^\dagger c_{n^\prime} | b_f \rangle \langle b_f | b_i\rangle 
\langle b_i | c_p^\dagger c_n | 0^+_i \rangle . \label{eq:M0v}
\end{eqnarray}
The initial and final states are denoted by $| 0^+_i\rangle$ and $| 0^+_f\rangle$, respectively, and 
$|b_{i \textrm{\ or\ }f}\rangle$ is the intermediate state obtained by the proton-neutron QRPA, e.g.~\cite{Mut89}\footnote{For simplicity, I call this theory QRPA in this paper.}. Since those states depend on the ground states, two complete sets of the QRPA solutions are used. 
The proton (neutron) is denoted by $p$ ($n$), and $c^\dagger_p$ ($c_p)$ stands for the creation (annihilation) operator of the specified particle. $V(r_{12},\bar{E}_b)$ is the two-body transition operator of the neutrinoless double-$\beta$  decay including the neutrino potential \cite{Doi85} ($r_{12}$ is the two-nucleon distance). $V(r_{12},\bar{E}_b)$ used in my calculation consists of the double Gamow-Teller and double Fermi operators. $\bar{E}_b$ is the average energy of the intermediate states relevant to $M^{(0\nu)}$.  

\section{Charge-Change Strength Function}

\subsection{Relation between Charge-Change Reaction and Double-$\bm{\beta}$ Decay}

The charge-change reaction is caused by the strong interaction. Therefore, the information obtained from this reaction is free from $g_A$, of which the effective value appropriate for the approximate calculations is not yet established. 

The Gamow-Teller strength function is obtained from the experimental charge-change cross section of the ($p$, $n$) and/or ($n$, $p$) reaction(s) through the impulse approximation and extrapolation of the cross section to the vanishing momentum transfer \cite{Yak09}. The charge-change reaction is enhanced in the measurement at or near the zero degree with the high incident energy.  The Gamow-Teller strength function is defined by
\begin{eqnarray}
\langle b | \bm{\sigma} \bm{\tau}
 | 0^+\rangle = \sum_{np} \langle p | \bm{\sigma} \bm{\tau} | n \rangle 
\langle b | c^\dagger_p c_n | 0^+\rangle. \label{eq:GTsf}
\end{eqnarray}
As seen from this equation and Eq.~(\ref{eq:M0v}), the transition density $\langle b | c^\dagger_p c_n | 0^+ \rangle$ is shared by those transition matrix elements, if the initial and final states of the charge-change reactions are chosen suitably to the neutrinoless double-$\beta$ decay. 
The cross sections of $^{48}$Ca($p$,$n$)$^{48}$Sc and $^{48}$Ti($n$,$p$)$^{48}$Sc have been measured \cite{Yak09}. These reactions fit to the neutrinoless double-$\beta$ decay of $^{48}$Ca  $\rightarrow$ $^{48}$Ti in the above sense. Note that the $J^\pi$ of the final states of the Gamow-Teller transition is limited to $1^+$, meanwhile there is no constraint on the $J^\pi$ of the intermediate states of the neutrinoless double-$\beta$ decay because of the neutrino potential. 

\subsection{Implication of Experimental Charge-Change Strength Function}
\begin{figure}[t]
\centering
\includegraphics[scale=0.65]{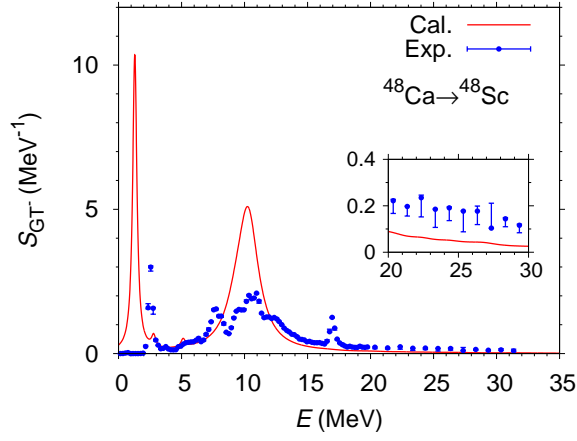}
  \caption[]{Experimental \cite{Yak09} (symbols) and calculated \cite{Ter18} (solid line) Gamow-Teller ($\beta$-decay like) strength functions of $^{48}$Ca($p$, $n$)$^{48}$Sc. The inset is a magnified figure of the high-energy region.}\label{fig:strfn_gt}
\end{figure}

\begin{figure}[t]
\centering
\includegraphics[scale=0.65]{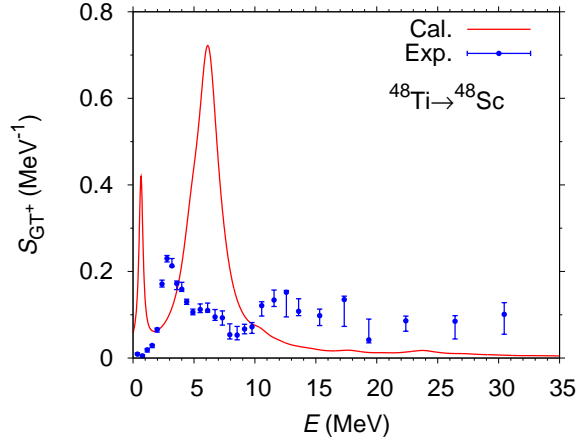}
  \caption[]{Experimental \cite{Yak09} (symbols) and calculated \cite{Ter18} (solid line) Gamow-Teller ($\beta^+$-decay like) strength functions of $^{48}$Ti($n$, $p$)$^{48}$Sc.}\label{fig:strfn_gt_p2n}
\end{figure}

The QRPA calculation was performed with the Skyrme (SkM$^\ast$ \cite{Bar82}) and contact volume-type pairing interactions. The same interactions were used for the Hartree-Fock-Bogoliubov (HFB) \cite{Ter03,Obe03,Bla05,Obe07} calculations prior to the QRPA calculations. Figure \ref{fig:strfn_gt} shows the calculated and experimental Gamow-Teller ($\beta$-decay like) transition, and Figure \ref{fig:strfn_gt_p2n} illustrates the analogous transition ($\beta^+$-decay like) of $^{48}$Ti. 


I stress that the calculated strength functions satisfy the sum rule of the Gamow-Teller transitions 
\begin{eqnarray}
&&{\cal S}^\textrm{\scriptsize QRPA}_{\beta-} - {\cal S}^\textrm{\scriptsize QRPA}_{\beta+} = 24.638 - 0.633 = 24.005, (^{48}\textrm{Ca}), \nonumber \\
&&{\cal S}^\textrm{\scriptsize QRPA}_{\beta-} - {\cal S}^\textrm{\scriptsize QRPA}_{\beta+} = 15.257 - 3.268 = 11.989, (^{48}\textrm{Ti}). \nonumber
\end{eqnarray}
 The two terms of the upper equation stand for the summation of the Gamow-Teller transition strengths of the $\beta$-decay-like ($^{48}$Ca $\rightarrow$ $^{48}$Sc) and $\beta^+$-decay-like ($^{48}$Ca $\rightarrow$ $^{48}$K) transitions, and the two terms of the lower equation are those of $^{48}$Ti $\rightarrow$ $^{48}$V and $^{48}$Ti $\rightarrow$ $^{48}$Sc. The exact values are 24 ($^{48}$Ca) and 12 ($^{48}$Ti). The sum of the measured transition strengths of $^{48}$Ca is 64$\pm$9 \% of the sum-rule value. Since the contribution of the $\beta^+$-decay-like transition is negative (not measured in this experiment), the data cannot satisfy the sum-rule. Note also that the tail of the experimental strength function of $^{48}$Ca is decreasing. If the measurement is extended to a higher-energy region, the sum rule would not be satisfied. 

There is another major discrepancy (other than the peak energies) between the experimental and calculated transition-strength distributions.  The calculated strength function is larger than the experimental one in $E < 12$ MeV ($^{48}$Ca $\rightarrow$ $^{48}$Sc) and $E<10$ MeV ($^{48}$Ti $\rightarrow$ $^{48}$Sc). In the higher-energy region, this relation is inverted. 

\begin{figure}[t]
\centering
\includegraphics[scale=0.65]{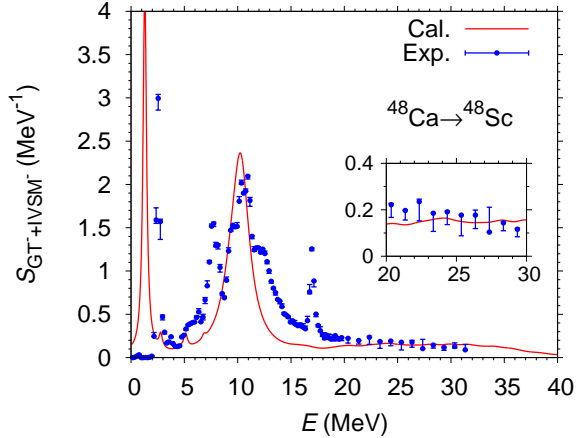}
  \caption[]{Experimental \cite{Yak09} (symbols) and modified \cite{Ter18} (solid line) charge-change ($\beta$-decay like) strength functions of $^{48}$Ca($p$, $n$)$^{48}$Sc. The inset is a magnified figure of the high-energy region.}\label{fig:strfn_mix}
\end{figure}

\begin{figure}[t]
\centering
\includegraphics[scale=0.65]{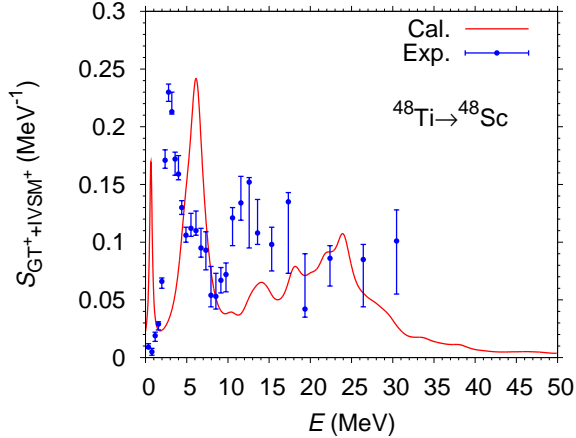}
  \caption[]{Experimental \cite{Yak09} (symbols) and modified \cite{Ter18} (solid line) charge-change ($\beta^+$-decay like) strength functions of $^{48}$Ti($n$, $p$)$^{48}$Sc.}\label{fig:strfn_mix_p2n}
\end{figure}

The reasonable explanation of the experimental data is to add another term to the transition operator;
\begin{eqnarray}
\bm{\sigma}\bm{\tau} + \alpha r^2 \bm{\sigma}\bm{\tau}, 
\end{eqnarray} 
where the constant $\alpha$ is determined so as to satisfy the height of the experimental strength function in the higher-energy region on average. I used $\alpha=-0.03$ fm$^{-2}$ ($^{48}$Ca $\rightarrow$ $^{48}$Sc) and $-0.0253$ fm$^{-2}$ ($^{48}$Ti $\rightarrow$ $^{48}$Sc), and the results of Figs.~\ref{fig:strfn_mix} and \ref{fig:strfn_mix_p2n} were obtained. The experimental data in the lower-energy region are well reproduced; this is the non-trivial achievement of the new idea. The choice of the measurement angle and the incident energy does not constrain the detail of the transition operator. Thus, the data imply that the transition operator includes the $r$-dependence (that second term is called the isovector spin monopole operator \cite{Ham00}). My transition density is confirmed indirectly by this reproduction of the data. 


The analysis of Ref.~\cite{Ter18} indicates that the mechanism of the improvement is explained by the decomposition of the transition operator
\begin{eqnarray}
\bm{\sigma}\bm{\tau} + \alpha r^2 \bm{\sigma}\bm{\tau} 
 = \left\{ 1 + \alpha\langle r^2\rangle_\mathrm{n1f7/2}
+ \alpha (r^2 - \langle r^2\rangle_\mathrm{n1f7/2}) \right\} \bm{\sigma}\bm{\tau}. 
\end{eqnarray}
The $\langle r^2 \rangle_\mathrm{n1f7/2}$ is the single-particle expectation value with respect to the excessive neutron\footnote{1f$_{7/2}$ indicates the first f$_{7/2}$.} compared to the protons in the ground state of $^{48}$Ca. The coordinate operator 
$\alpha(r^2 - \langle r^2 \rangle_\mathrm{n1f7/2})$ causes the so-called two-$\hbar \omega$ jump. The zero-$\hbar\omega$ component is subtracted by the constant because the transition matrix element of the wave functions in the coordinate space between the n1f$_{7/2}$ and p1f$_{7/2}$ is equal to the expectation value under the isospin symmetry \cite{Ham00}. The constant 
$(1 + \alpha\langle r^2 \rangle_\mathrm{n1f7/2})$ is an operator of the zero-$\hbar \omega$ jump. Therefore, the strength function in the lower-energy region is decreased, and that in the higher-energy region is increased by that modified transition operator.

\section{Reduced Half-Life}
I discuss the reduced half-life $R^{(0\nu)}_{1/2}$ \cite{Eng17} defined by
\begin{eqnarray}
T^{(0\nu)}_{1/2} = \frac{ R^{(0\nu)}_{1/2} }{\langle m_\nu \rangle^2}.  
\end{eqnarray}
As seen from this equation, $R^{(0\nu)}_{1/2}$ is the theoretical quantity necessary for determining $\langle m_\nu \rangle$ and also a substitute of $T^{(0\nu)}_{1/2}$ not yet obtained for discussion. If different  approximate calculations are correct, the $R^{(0\nu)}_{1/2}$ is unique because the true $T^{(0\nu)}_{1/2}$ and $\langle m_\nu \rangle$ are unique. Therefore, $R^{(\nu)}_{1/2}$ is useful for comparison of the calculations with different $g_A$. This feature is particularly useful for my calculation because my values of the effective $g_A$ are relatively small ($\sim$ 0.4$-$0.5) \cite{Ter18} as a result of the new method to determine the strength of the isoscalar pairing interaction, with which the HFB ground state is not near the instability to the QRPA excitations. Values between 1.27 and 1.0 are used for $g_A$ in many calculations of other groups; see e.g.~\cite{Eng17} and references therein. 
$R^{(0\nu)}_{1/2}$ is calculated by 
\begin{eqnarray}
R^{(0\nu)}_{1/2} = \frac{ m_e^2 }{ g_A^4 G^{(0\nu)} \left| M^{(0\nu)}\right|^2 }, 
\end{eqnarray}
[see Eq.~(\ref{eq:transition_probability_0vbb})]. 

\begin{figure}[t]
\centering
\includegraphics[scale=0.62]{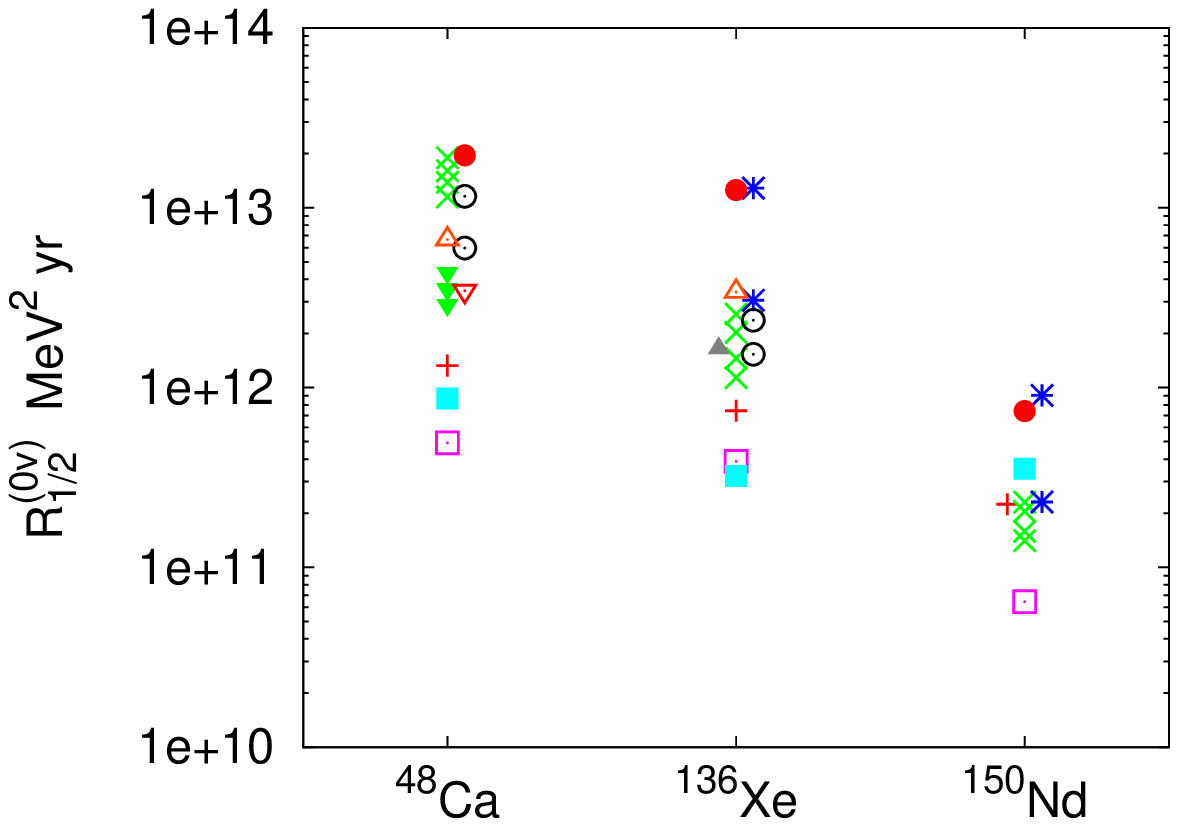}
  \caption[]{$R^{(0\nu)}_{1/2}$ calculated by different methods and groups for $^{48}$Ca, $^{136}$Xe, and $^{150}$Nd. See Fig.~\ref{fig:legend} for the legends.}\label{fig:R}
\centering
\includegraphics[scale=0.62]{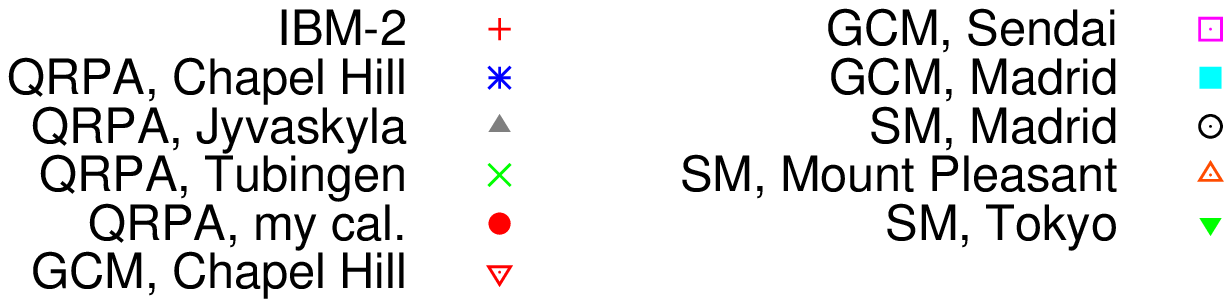}
 \vspace*{-30pt} 
 \caption[]{ Legends of Fig.~\ref{fig:R}. The references are as follows: $^{48}$Ca, \cite{Sim13} (QRPA T\"{u}bingen); \cite{Hor16} (SM, Mount Pleasant); \cite{Iwa16} (SM, Tokyo); \cite{Bar15} (IBM-2); \cite{Vaq13} (GCM, Madrid); \cite{Yao15} (GCM, Sendai); \cite{Men09} (SM, Madrid); \cite{Jia17} (GCM, Chapel Hill); \cite{Ter18} (QRPA, my calculation). $^{136}$Xe, \cite{Bar15} (IBM-2); \cite{Sim13} (QRPA, T\"{u}bingen); \cite{Mus13} (QRPA, Chapel Hill); \cite{Yao15} (GCM, Sendai); \cite{Vaq13} (GCM, Madrid); \cite{Men09} (SM, Madrid); \cite{Hor16} (SM, Mount Pleasant); \cite{Hyv15} (QRPA, Jyvaskyla);   current paper (QRPA, my calculation). $^{150}$Nd, \cite{Bar15} (IBM-1); \cite{Fan15} (QRPA, T\"{u}bingen); \cite{Mus13} (QRPA, Chapel Hill); \cite{Yao15} (GCM, Sendai); \cite{Vaq13} (GCM, Madrid); \cite{Ter15,Ter16} (QRPA, my calculation). SM, GCM, and IBM stand for shell model, generator-coordinate method, and interacting-boson model, respectively. }\label{fig:legend}
\end{figure}

The calculated $R^{(0\nu)}_{1/2}$ are shown by Fig.~\ref{fig:R} for three mother nuclei together with those of other groups (the legends are indicated by Fig.~\ref{fig:legend}). 
My results are the largest values; the main reason is the smallness of $g_A$. Let us assume that $\langle m_\nu \rangle=10$ meV. With my $R^{(0\nu)}_{1/2}$, $T^{(0\nu)}$ is predicted to be 2$\times$10$^{29}$ years for $^{48}$Ca. One can compare this half-life to the estimated age of the universe of (12$-$14)$\times$10$^{9}$ years \cite{Ade16}. The neutrinoless double-$\beta$ decay is an extremely rare decay. 

\section{Summary}
I have calculated the charge-change strength functions of $^{48}$Ca and $^{48}$Ti for confirming the transition density for the calculation of the nuclear matrix elements of the double-$\beta$ decay. As a by-product, it has been found that the transition operator implied by the experimental charge-change strength functions is a summation of the usual Gamow-Teller and isovector spin monopole operators. The mechanism of the modification was also discussed on the basis of the configuration of $^{48}$Ca. My charge-change transition density has been confirmed indirectly by the reproduction of the experimental strength function. Considering the difficulty that the nuclear matrix element cannot be confirmed by experiment, this check is very important. 

Subsequently, I have shown the calculated reduced half-life rather than the nuclear matrix elements because my effective value of $g_A$ is small. My values of $R^{(0\nu)}_{1/2}$ are the largest in the calculations by different methods and groups for the three nuclei shown. For now there is no method to confirm those values. One thing for sure under this uncertainty is that in any way the half-life to the neutrinoless double-$\beta$ decay is longer by many orders of years than the estimated age of the universe. 
The task of the theorists is to accumulate the checks and improvements of the calculations of the nuclear matrix elements for improving the reliability. 

\section*{Acknowledgments}
The computation of this study has been performed by the K computer at RIKEN Center for Computational Science through the program of High Performance Computing Infrastructure (subject ID hp170288 and local account k03936) and Oakforest-PACS at Information Technology Center, University of Tokyo through the Multidisciplinary Cooperative Research Program of Center for Computational Sciences, University of Tsukuba (project ID xg18i006 and user account x10346). 
This study is supported by European Regional Development Fund, Project ``Engineering applications of microworld physics'' (No.~CZ.02.1.01/0.0/ 0.0/16\_019/0000766).

\end{document}